\documentclass[preprint]{elsarticle}
\usepackage{lineno,hyperref,pdflscape}
\modulolinenumbers[5]

\usepackage{multicol}
\usepackage{color}
\usepackage{xspace}
\usepackage{enumerate}
\usepackage{amsmath} 
\usepackage{makecell}
\usepackage{ulem} 
\usepackage{placeins}
\usepackage{threeparttable}

\usepackage{tabularx}
\usepackage{mathtools} 
\usepackage{amsfonts}
\newcolumntype{M}{>{\centering\arraybackslash}m{1.85cm}}
\usepackage[export]{adjustbox}
\usepackage{float}
\usepackage{braket}
\usepackage{lipsum}
\usepackage{longtable}
\usepackage{lscape}
\graphicspath{{figure/}}

\usepackage{booktabs}
\usepackage{blindtext}
\makeatletter
\newcommand{\colorcaption}[2][]{%
	\begingroup%
	\renewcommand{\@caption@fignum@sep}{ (Color online). }%
	\caption[#1]{#2}%
	\endgroup%
}

\makeatother

\usepackage{amssymb}
\usepackage{graphicx}
\usepackage{rotating}
\usepackage{tikz}
\begin{document}

	\begin{frontmatter}

\title{Nuclear structure properties of $^{184-194}$Pb isotopes and isomers}

\address[IITR]{Department of Physics, Indian Institute of Technology Roorkee- Roorkee 247667, India}

\author[IITR]{Sakshi Shukla}
\author[IITR]{Praveen C. Srivastava}


\begin{abstract}
In the present work, we study nuclear structure properties of the $^{184-194}$Pb isotopes within the framework of the nuclear shell-model. We have performed shell-model calculations using KHH7B and KHHE interactions. We have reported results for energy spectra, electromagnetic properties such as quadrupole moment ($Q$), magnetic moment ($\mu$), $B(E2)$, and $B(M1)$ transition strengths, and compared the shell-model results with the available experimental data. The shell-model results for the half-lives and seniority quantum numbers ($v$) are also reported for the isomeric states. 

\end{abstract}


\end{frontmatter}




\section{Introduction}
\label{introduction}

In recent years, there are numerous experimental activities going on for the study of atomic nuclei in the Pb region \cite{Plaza}. 
The nuclear shell-model (SM) is the fundamental microscopic method to understand the complex nuclear structure such as deformation \cite{Brown_PRL,Isacker,Rejmund,Frank,Yuan}.  Light Pb nuclei exhibit a wide range of coexisting nuclear shapes and exotic excitations due to the interplay between single-particle motion, collectivity, and pairing \cite{Isacker, Rejmund,Duppen, Heyde, Julin}.  For instance, the unique shape staggering discovered in the mercury isotopes \cite{Hg} that describes the coexistence of single-particle and collective degrees of freedom, the shape coexistence explored by $\alpha$-decay observed in nuclei around Pb \cite{Andreyev1} etc. The neutron-rich nuclei around {\color{black}$^{208}$Pb doubly closed magic nucleus with $Z$=82 and $N=126$} undergo $r$-process (rapid neutron
capture nucleosynthesis). Therefore, the study of nuclei around $N$ = $126$, helps us to better understand astrophysical $r$-process, which results the formation of heavier elements \cite{Nieto,r-process}.

The results of in-beam $\gamma$-ray
spectroscopy of the very neutron-deficient nuclear $^{184}$Pb, using the JUROGAM II spectrometer at the RITU separator are reported in Ref. \cite{Ojala}. In this work, first time non-yrast states have been observed below $N$ = $104$ midshell. The 
non-yrast structures of the neutron mid-shell nucleus $^{186}$Pb reported in Ref. \cite{Pakarinen_186Pb}, and the evidence for octupole and $\gamma$ - vibrational bands are also reported. In Ref. \cite{Dracoulis_188Pb}, several isomers have been identified. Two of the isomers i.e. $11^-$ (136 ns) and $12^+$ (38 ns) feed the $10^+$ state of the yrast sequence. In Ref. \cite{Plaza}, the three different deformations near the ground state in $^{190}$Pb are reported.  
The shape of the ground state in $^{190,192}$Pb is discussed in Ref. \cite{shape1}, where the shape of the ground state of both these isotopes are found to be spherical by studying $\beta$-decay of $^{190,192}$Pb using total absorption technique at the ISOLDE (CERN) facility. The experimental measurement of $E3$ transition strength for $11^-$ to $8^+$ transitions for $^{194,196}$Pb isotopes are reported in Ref. \cite{isomer4}, this is due to change of configurations from $\pi[h_{9/2}i_{13/2}]$ ($11^-$) to
$[h^2_{9/2}]$ ($8^+$).

  In this region, multiparticle-multihole (np-nh) proton excitations across the closed Z = 82 shell generally leads to the occurrence of excited $0^+$ states \cite{Heyde}. The low-lying $0^+$ states form the basis for the rotational bands observed in the even-mass Pb isotopes \cite{Rahkila,Jenkins,Pakarinen1,Dracoulis1,Dracoulis2}. When approaching the neutron mid-shell at N = 104, they intrude down around the energies of the spherical ground states \cite{Andreyev1, Andreyev2}. Several experimental data are available for the extended series of even-even Pb nuclei near $^{208}_{82}$Pb 
  \cite{broda, Wahid, Barzakh,  podolyak} for the study of nuclear structure and collectivity. In addition, different types of correlations such as pairing, quadrupole, and octupole are discussed in Refs. \cite{Isacker,Rejmund,octupole}. 
Several isomeric states are known in this region \cite{Isomer1, Isomer2, Isomer3,Isomer4}. 

In the present work, we have done a systematic shell-model study of $^{184-194}$Pb isotopes to investigate nuclear structure properties and isomeric states. 
Earlier in this series, the low-lying states in the Pb isotopes for the mass range A = 196-206 are explored in Ref. \cite{Sakshi1} and the mass range A = 211-215 is explored in Ref. \cite{Zamick}.
Our group has also studied $^{204-210}$Tl \cite{Bharti_Tl}, $^{200-210}$Po \cite{Sakshi2}, $^{204-213}$Bi \cite{Sakshi3}, $^{206-216}$Rn \cite{Bharti_Rn} nuclei around $^{208}$Pb, and $N$ = 126 isotones \cite{Bhoy_Sn_Pb}.


This paper is arranged as follows: In section \ref{sec2}, theoretical formalism of the interactions used in our calculations is discussed, section \ref{sec3} belongs to the results obtained from the interactions used, and section \ref{sec4} comprises the key findings and conclusions.

\section{Theoretical Framework}\label{sec2}

The nuclear shell-model Hamiltonian can be expressed in terms of
single-particle energies and two-nucleon interactions,
\begin{equation}
H = \sum _{\alpha}\epsilon _{\alpha}c_{\alpha}^{\dagger}c_{\alpha}+
\frac{1}{4}\sum _{\alpha \beta \gamma \delta JT}\langle j_{\alpha }j_{ \beta
}|V|j_{\gamma }j_{\delta} \rangle _{JT}c_{\alpha}^{\dagger}c_{
\beta}^{\dagger}c_{\delta }c_{\gamma },\label{eq1}
\end{equation}
where $\alpha =\{n,l,j,t\}$ stand for the single-particle orbitals and
$\epsilon _{\alpha}$ are corresponding single-particle energies. 
The $c_{\alpha}^{\dagger}$ and $c_{\alpha}$ are the fermion creation and annihilation operators, respectively. The antisymmetrized two-body matrix elements coupled to spin
$J$ and isospin $T$ is given by $\langle j_{\alpha }j_{\beta }|V|j_{\gamma }j_{\delta }\rangle
_{JT}$. To diagonalize the matrices, the shell-model code KSHELL \cite{KSHELL} has been employed. 
 To study nuclear structure properties of $^{184-194}$Pb isotopes, we have performed shell-model calculations using KHH7B interaction \cite{KHH7B1,KHH7B2,KHH7B3}, in addition we have also done calculations using KHHE interaction \cite{KHH7B1,KHH7B2,KHH7B3}. The KHH7B interaction consists of total 14 orbitals out of which there are seven proton orbitals between $Z = 58-114$: $1d_{5/2}$, $1d_{3/2}$, $2s_{1/2}$, $0h_{11/2}$, $0h_{9/2}$, $1f_{7/2}$, $0i_{13/2}$ and seven neutron orbitals between $N = 100-164$: $1f_{5/2}$, $2p_{3/2}$, $2p_{1/2}$, $0i_{13/2}$, $1g_{9/2}$, $0i_{11/2}$, $0j_{15/2}$. We have completely filled the proton orbital below Z=82. {\color{black}Here, we have not included proton excitations across $Z$ = 82 in our calculations.} For valence neutrons, restrictions are such that only orbitals below N=126 are allowed.
 The model space of KHHE interaction consists of a total of 11 orbitals, among which $0g_{7/2}$, $1d_{5/2}$, $1d_{3/2}$, $2s_{1/2}$, and $0h_{11/2}$ are proton orbitals and $0h_{9/2}$, $1f_{7/2}$, $1f_{5/2}$, $2p_{3/2}$, $2p_{1/2}$, and $0i_{13/2}$ are neutron orbitals. The holes in a $^{208}$Pb core served as the foundation for the KHHE interaction. For this interaction, we have completely filled all proton orbitals, and to make the calculation feasible, we have restricted the valence neutrons orbitals with $\nu$($h_{9/2}^{8-10}$$f_{7/2}^{6-8}$$f_{5/2}^{0-6}$$p_{3/2}^{0-4}$$p_{1/2}^{0-2}$$i_{13/2}^{0-14}$) partition. 
The shell-model results are shown in the Figs. 1-12  and Tables 1-6.

\section{Results and Discussions}\label{sec3}


\subsubsection{$^{184}$Pb}
\begin{figure}
\begin{center}
\includegraphics[width=9.00cm,height=9cm]{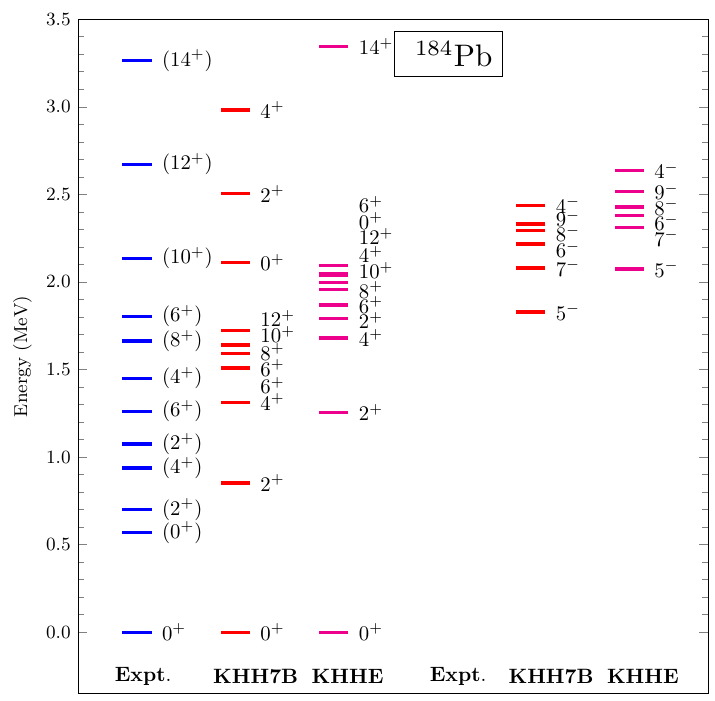}
\caption{\label{184Pb}Comparison between calculated and experimental \cite{NNDC} energy levels for $^{184}$Pb isotope.}
\end{center}
\end{figure}
The results of our calculations for the $^{184}$Pb isotope are shown in comparison with the experimental data in Fig. \ref{184Pb}. 
The configurations of selected states are shown in the Table \ref{configuration}.
Here, we have taken experimental energy states up to 3.3 MeV excitation energy. 
There is no experimental data for negative parity state is available, though SM predicts $5^-_1$ to be lowest lying negative parity state. Results obtained from KHH7B interaction are slightly lower than that of the KHHE interaction. We see the highest discrepancy in excitation energy for $0^+_2$ state in comparison to the experimental state from both the interactions. The second $0^+$ state comes due to two neutrons in $p_{3/2}$ orbital using KHH7B interaction, because the $p_{3/2}$ orbital is located distant from the Fermi surface, therefore the excitation energy of $0^+_2$ state is very high. 
First $0^+$ state is obtained from configuration $\nu( i_{13/2}^2$), whereas second $0^+$ state is obtained due to $\nu( p_{3/2}^2$) from KHH7B interaction.
The states $5^-_1$, $6^-_1$, and $7^-_1$ are obtained from coupling of  $\nu (p_{3/2}^1)$ with $ \nu (i_{13/2}^1)$ orbitals, whereas, $4^-_1$ state is obtained due to $\nu (f_{5/2}^1i_{13/2}^1)$. The $i_{13/2}$ orbital lies lowest from the Fermi surface, therefore $4^-_1$ state has a large excitation energy. 


\subsubsection{$^{186}$Pb}
\begin{figure}
\begin{center}
\includegraphics[width=9.00cm,height=9cm]{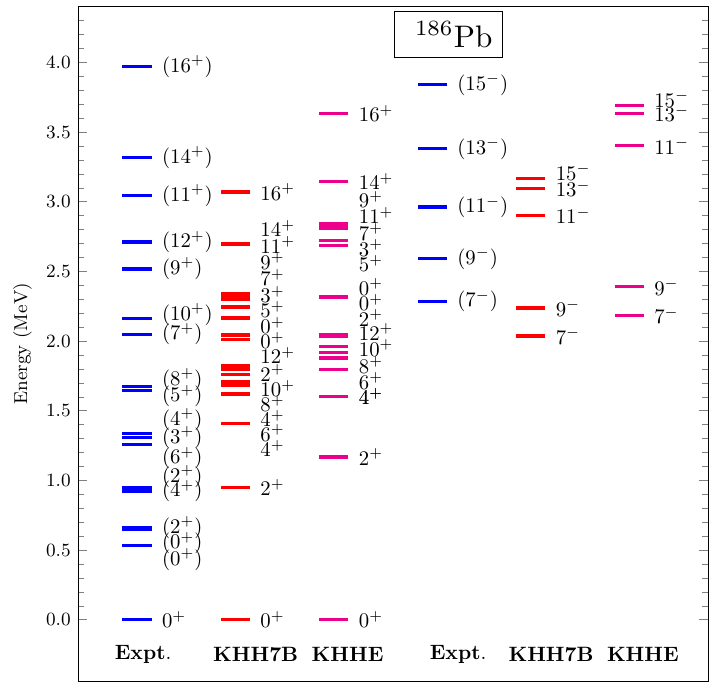}
\caption{\label{186Pb}Comparison between calculated and experimental \cite{NNDC} energy levels for $^{186}$Pb isotope.}
\end{center}
\end{figure}

The results of our calculations for the $^{186}$Pb isotope are shown in comparison with the experimental data in Fig. \ref{186Pb}. Here, we have taken experimental energy states up to 4.0 MeV excitation energy. The lowest-lying experimental negative parity state is also reproduced by SM calculation. Except the ground state, all the other states are tentatively assigned in this isotope. Negative parity states are reproduced in the correct order as the experimental data, whereas in most of the cases the results are underestimated by both the interactions, although the results from KHHE interaction are closer to the experimental value. All the states reported here using KHH7B interaction are obtained due to $\nu(i_{13/2}^4)$. From KHHE interaction yrast even $0^+-12^+$ states are also coming from $\nu(i_{13/2}^4)$, whereas $0^+_2$ state is coming from $\nu(p_{3/2}^2i_{13/2}^2)$. The $0^+_{g.s.}$, $0^+_2$ and $0^+_3$ states in $^{186}$Pb are spherical, prolate and oblate in shape, respectively \cite{186Pb_Shape}.
Experimentally, the yrast $E2$ cascade between $14^+\rightarrow12^+\rightarrow10^+\rightarrow8^+\rightarrow6^+\rightarrow4^+\rightarrow2^+$ states are observed \cite{186Pb_Shape}, among which experimental data for $B(E2)$ transition strength are available up to yrast $8^+$ state and corresponding theoretical value is also calculated (shown in Table \ref{be2}). 


\subsubsection{$^{188}$Pb}
\begin{figure}
\begin{center}
\includegraphics[width=9.00cm,height=9cm]{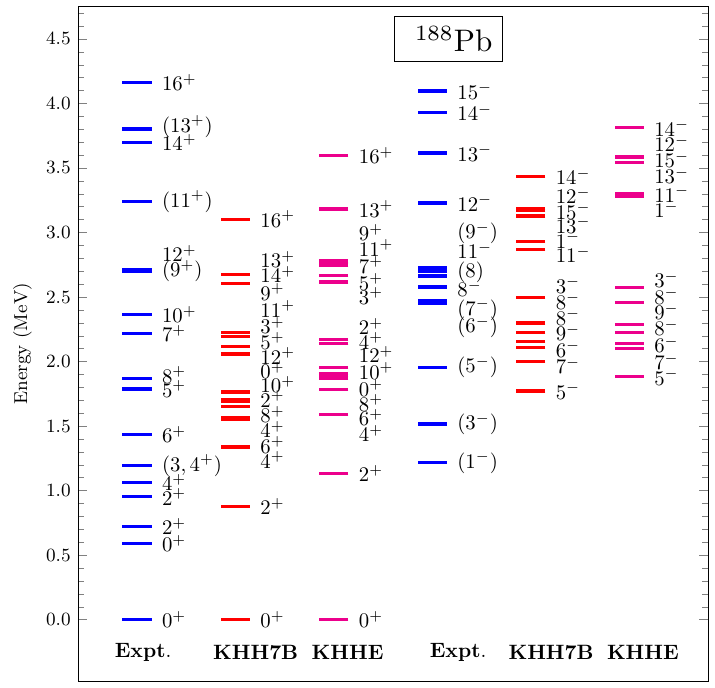}
\caption{\label{188Pb}Comparison between calculated and experimental \cite{NNDC} energy levels for $^{188}$Pb isotope.}
\end{center}
\end{figure}

The results of our calculations for the $^{188}$Pb isotope are shown in comparison with the experimental data in Fig. \ref{188Pb}.  Here, we have taken experimental energy states up to 4.2 MeV excitation energy. We are unable to reproduce lowest-lying experimentally observed negative parity state, due to our restricted calculation. The calculated magnetic moment for the $11^-_1$ state is having negative sign, when we consider its configuration which is $\nu(p_{3/2}^1i_{13/2}^5)$, we found that the magnetic moment should be negative for the $11^-_1$ state, whereas its corresponding experimental value has positive sign (shown in Table \ref{qm}). 
Experimentally, $(7^-)$ state at 2.474 MeV excitation energy has tentative assignment. It is observed experimentally that $8^-$ state at 2.577 MeV excitation energy decays to $(7^-)$ state by $M1$ transition with the transition strength 9.1$\times$10$^{-7}$ (15) W.u., although it is much smaller than the SM calculated $B(M1;8^-_1\rightarrow7^-_1)$ value as shown in Table \ref{bm1}.


\subsubsection{$^{190}$Pb}
\begin{figure}
\begin{center}
\includegraphics[width=9.00cm,height=9cm]{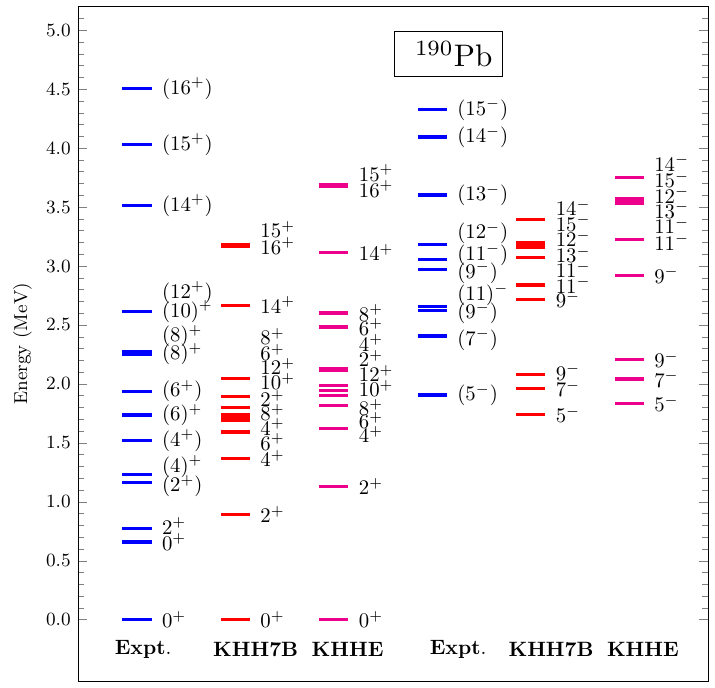}
\caption{\label{190Pb}Comparison between calculated and experimental \cite{NNDC} energy levels for $^{190}$Pb isotope.}
\end{center}
\end{figure}
The results of our calculations for the $^{190}$Pb isotope are shown in comparison with the experimental data in Fig. \ref{190Pb}.  Here, we have taken experimental energy states up to 4.5 MeV excitation energy. In this isotope, all the experimentally observed states are tentative except the ground and first two excited states. We are able to reproduce experimentally observed ground state and lowest-lying negative parity state by both interactions. All the even yrast $0^+-16^+$ states reported here are coming from $\nu(i_{13/2}^8)$ configuration.
The yrast $5^-$, $7^-$, $12^-$, $13^-$, $14^-$, and $15^-$ are coming from $\nu(p_{3/2}^1i_{13/2}^7)$ configuration, whereas yrast $9^-$, and $11^-$ states are coming from $\nu(f_{5/2}^1i_{13/2}^7)$ from both the interactions. In the Ref.\cite{Dracoulis2} the configuration of $11^-_1$ state is $\pi(i_{13/2}h_{9/2})$ i.e. two proton intruder state, whereas $9^-_1$ share a similar configuration as obtained in our calculation. For this isotope experimental data for only $B(E2;10^+_1\rightarrow8_2^+)$ value is available, which is equal to 0.012 W.u., our corresponding SM obtained value is 0.8 W.u. There are no experimental data available for the magnetic and quadrupole moment, and $M1$ transition strength, although we have reported SM obtained results for some states in Table \ref{bm1}, and  \ref{qm}, respectively. This may be useful for comparison with the upcoming experimental data.

\subsubsection{$^{192}$Pb}
\begin{figure}
\begin{center}
\includegraphics[width=9.00cm,height=9cm]{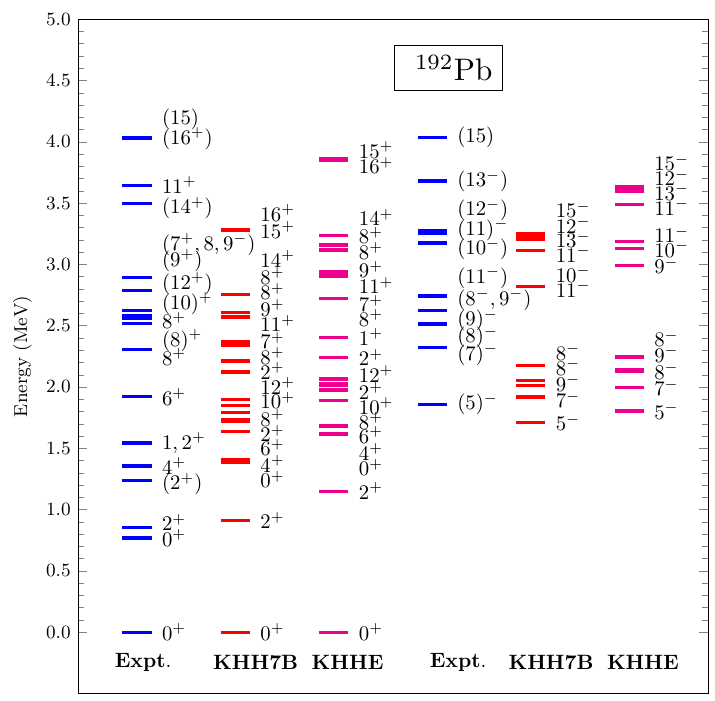}
\caption{\label{192Pb}Comparison between calculated and experimental \cite{NNDC} energy levels for $^{192}$Pb isotope.}
\end{center}
\end{figure}
The results of our calculations for the $^{192}$Pb isotope are shown in comparison with the experimental data in Fig. \ref{192Pb}.  Here, we have taken experimental energy states up to 4.04 MeV excitation energy. We are able to reproduce the lowest-lying negative parity state by using KHH7B interaction. Experimentally, the spins of all the negative parity states are tentative. If we compare these tentative experimental states with our calculated SM states, we find that our calculated results are under-estimated. Experimentally, at 4.035 MeV excitation energy, $(15)$ state with unconfirmed parity state, the calculated result of $15^+_1$ with KHHE
 interaction at 3.864 MeV is close to this state, therefore we can predict this state may belong to $15^+_1$ state.  
The calculated magnetic moment for the $12^+_1$ isomeric state is -3.51 $\mu_N$, which is 1.7 times greater than its experimental value i.e. -2.076 (24) $\mu_N$, whereas the calculated quadrupole moment for the $12^+_1$ state is 2.5 times smaller than its experimental value, as shown in Table \ref{qm}. 

\subsubsection{$^{194}$Pb}
\begin{figure}
\begin{center}
\includegraphics[width=9.00cm,height=9cm]{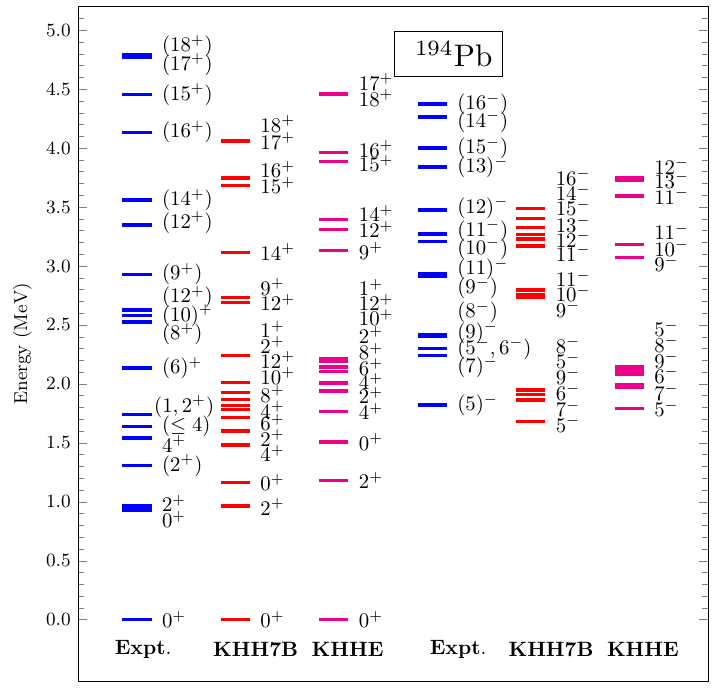}
\caption{\label{194Pb}Comparison between calculated and experimental \cite{NNDC} energy levels for $^{194}$Pb isotope.}
\end{center}
\end{figure}
The results of our calculations for the $^{194}$Pb isotope are shown in comparison with the experimental data in Fig. \ref{194Pb}.  Here, we have taken experimental energy states up to 4.8 MeV excitation energy. Experimentally, except the yrast $0^+$, $2^+$, $4^+$, and second yrast $0^+$ states, all the other states are tentative. Yrast even $0^+-12^+$ states obtained from $\nu(i_{13/2}^{12})$ configurations using KHH7B interaction, whereas yrast $9^+$, $14^+$, $16^+$, $17^+$, and $18^+$ states are obtained from $\nu(p_{3/2}^2i_{13/2}^{10})$ configuration. The $1^+$, and $15^+$ states are coming from $\nu(f_{5/2}^1p_{3/2}^1i_{13/2}^{10})$ configuration. The negative parity states $5^-$, $14^-$, $15^-$, and $16^-$ are obtained from $\nu(p_{3/2}^1i_{13/2}^{11})$, whereas rest of the negative parity states reported here are obtained from $\nu(f_{5/2}^1i_{13/2}^{11})$ configurations. The quadrupole moment for the $11^-_1$ [$\nu(f_{5/2}^1i_{13/2}^{11})$] isomeric state is obtained to be -0.07 eb from the KHH7B interaction, whereas its experimental value is 3.6 (4) eb \cite{quadra_194Pb} having opposite sign with respect to theoretically calculated value, when we calculate Q($11^-_1$) using single particle estimate for effective charges $e_p,e_n$=(1.5,0.5)e, we found that the $Q_{\text{total}}$ $(e_{eff}^nQ(\nu f_{5/2}^{1})+e_{eff}^nQ(\nu i_{13/2}^{11})$) is negative for the given configuration, therefore our SM result is also supported by single-particle estimate calculation for quadrupole moment. The decay of $8^+_4$ is observed experimentally and found to decay in $6^+_4$, $8^+_1$, and $9^+_1$ states via $E2$, $M1+E2$, and $M1$ transitions, respectively. We have calculated $B(M1)$, and $B(E2)$ values for these transitions, which is reported in Table \ref{be2}, and \ref{bm1}, we found that both the theoretical and experimental observation corresponding to $B(M1;8^+_4\rightarrow9^+_1)$, and $B(M1;8^+_4\rightarrow8^+_1)$ provide smaller transition strengths. Which suggests $B(M1;8^+_4\rightarrow9^+_1)$, and $B(M1;8^+_4\rightarrow8^+_1)$ are not the primary transitions, whereas experimental $B(E2;8^+_4\rightarrow6^+_4)$ value is very high suggesting it to be primary transition, though theoretically we are unable to reproduce experimental $B(E2)$ value. Experimentally mixing ratio for the $(8^-)$, and $(7)^-$ states are obtained to be $<$0.7, if we compare $(8^-)$ and $(7)^-$ states with the theoretically calculated $8^-_1$ and $7^-_1$, though our calculated results are under predicted, then the calculated mixing ratio for the $8^- \rightarrow 7^-$ transition is -0.64, although there is a change in sign.
With a decreasing neutron number, the excitation energy of the second 0$^+$ state rapidly drops in the lighter Pb isotopes.  In $^{184-194}$Pb, it even turns into the first excited state. These low-lying $0^+$ states could be understood in the context of a SM as coexisting deformed states that are brought about by proton pair excitations across the $Z$ = 82 shell gap.
As the neutron number approaches mid-shell, the enlarged valence-neutron space and enhanced residual correlations lower the energy of these core-excited configurations, making the intruder $0^+$ states energetically favorable.
Whereas, in our case $2^+_1$ state is obtained as the first excited state in $^{184-194}$Pb isotope. This remains a challenge for existing interactions.

\begin{table}
\centering
\caption{\label{configuration} The calculated shell-model wave functions corresponding to different excited states in Pb isotopes with KHH7B effective interaction.}
\vspace*{5mm}

\begin{tabular}{cccc}
\hline 
      Nucleus & J$^\pi$ & Wavefunction & Probability(\%)\\
      \hline

 \hline
      $^{184}$Pb & 0$^+$\hspace{0.5cm} &  $\nu (f_{5/2}^0p_{3/2}^0p_{1/2}^0i_{13/2}^{2})$ & 83.54\\
 & 2$^+$\hspace{0.5cm} &  $\nu (f_{5/2}^0p_{3/2}^0p_{1/2}^0i_{13/2}^{2})$ & 96.48\\ 
 & 4$^+$\hspace{0.5cm} &  $\nu (f_{5/2}^0p_{3/2}^0p_{1/2}^0i_{13/2}^{2})$ & 98.87\\ 
 & 6$^+$\hspace{0.5cm} &  $\nu (f_{5/2}^0p_{3/2}^0p_{1/2}^0i_{13/2}^{2})$ & 100\\ 
  & 8$^+$\hspace{0.5cm} &  $\nu (f_{5/2}^0p_{3/2}^0p_{1/2}^0i_{13/2}^{2})$ & 100\\ 

 \hline
      $^{186}$Pb & 0$^+$\hspace{0.5cm} & $\nu (f_{5/2}^0p_{3/2}^0p_{1/2}^0i_{13/2}^{4})$ & 69.04\\
       $ $ & 2$^+$\hspace{0.5cm} & $\nu (f_{5/2}^0p_{3/2}^0p_{1/2}^0i_{13/2}^{4})$ & 79.08\\
          $ $ & 4$^+$\hspace{0.5cm} & $\nu (f_{5/2}^0p_{3/2}^0p_{1/2}^0i_{13/2}^{4})$ &83.32\\
             $ $ & 6$^+$\hspace{0.5cm} & $\nu (f_{5/2}^0p_{3/2}^0p_{1/2}^0i_{13/2}^{4})$ &84.48\\
               $ $ & 8$^+$\hspace{0.5cm} & $\nu (f_{5/2}^0p_{3/2}^0p_{1/2}^0i_{13/2}^{4})$ &84.37\\
       $ $ & 7$^-$\hspace{0.5cm} & $\nu (f_{5/2}^0p_{3/2}^1p_{1/2}^0i_{13/2}^{3})$   &57.93\\
        $ $ & 9$^-$\hspace{0.5cm} & $\nu (f_{5/2}^1p_{3/2}^1p_{1/2}^0i_{13/2}^{3})$   &85.21\\
         $ $ & 11$^-$\hspace{0.5cm} & $\nu (f_{5/2}^1p_{3/2}^0p_{1/2}^0i_{13/2}^{3})$   &52.99\\
   \hline

      $^{188}$Pb    & 0$^+$\hspace{0.5cm} & $\nu (f_{5/2}^0p_{3/2}^0p_{1/2}^0i_{13/2}^{6})$  & 56.36\\
      $ $ & 2$^+$\hspace{0.5cm} & $\nu (f_{5/2}^0p_{3/2}^0p_{1/2}^0i_{13/2}^{6})$   &63.51\\
    $ $ & 4$^+$\hspace{0.5cm} & $\nu (f_{5/2}^0p_{3/2}^0p_{1/2}^0i_{13/2}^{6})$   &69.66\\      
  $ $ & 6$^+$\hspace{0.5cm} & $\nu (f_{5/2}^0p_{3/2}^0p_{1/2}^0i_{13/2}^{6})$   &70.94\\
    $ $ & 8$^+$\hspace{0.5cm} & $\nu (f_{5/2}^0p_{3/2}^0p_{1/2}^0i_{13/2}^{6})$   &70.70\\
    $ $ & 1$^-$\hspace{0.5cm} & $\nu (f_{5/2}^0p_{3/2}^1p_{1/2}^0i_{13/2}^{5})$   &65.79\\
     $ $ & 3$^-$\hspace{0.5cm} & $\nu (f_{5/2}^0p_{3/2}^1p_{1/2}^0i_{13/2}^{5})$   &42.85\\
 $ $ & 5$^-$\hspace{0.5cm} & $\nu (f_{5/2}^0p_{3/2}^1p_{1/2}^0i_{13/2}^{5})$   &70.45\\
  $ $ & 9$^-$\hspace{0.5cm} & $\nu (f_{5/2}^1p_{3/2}^0p_{1/2}^0i_{13/2}^{5})$   &73.74\\
  
    \hline
      $^{190}$Pb & 0$^+$\hspace{0.5cm} & $\nu (f_{5/2}^0p_{3/2}^0p_{1/2}^0i_{13/2}^{8})$ & 45.32\\
       $ $ & 2$^+$\hspace{0.5cm} & $\nu (f_{5/2}^0p_{3/2}^0p_{1/2}^0i_{13/2}^{8})$& 49.74\\
        $ $ & 4$^+$\hspace{0.5cm} & $\nu (f_{5/2}^0p_{3/2}^0p_{1/2}^0i_{13/2}^{8})$& 57.69\\
           $ $ & 6$^+$\hspace{0.5cm} & $\nu (f_{5/2}^0p_{3/2}^0p_{1/2}^0i_{13/2}^{8})$& 59.21\\
              $ $ & 8$^+$\hspace{0.5cm} & $\nu (f_{5/2}^0p_{3/2}^0p_{1/2}^0i_{13/2}^{8})$& 58.95\\
     $ $ & 5$^-$\hspace{0.5cm} & $\nu (f_{5/2}^0p_{3/2}^1p_{1/2}^0i_{13/2}^{7})$&60.37\\
     $ $ & 7$^-$\hspace{0.5cm} & $\nu (f_{5/2}^0p_{3/2}^1p_{1/2}^0i_{13/2}^{7})$&42.40\\
     $ $ & 9$^-$\hspace{0.5cm} & $\nu (f_{5/2}^0p_{3/2}^1p_{1/2}^0i_{13/2}^{7})$&63.81\\
       
   \hline


%


$^{192}$Pb &{0}$^+$\hspace{0.5cm} & $\nu (f_{5/2}^0p_{3/2}^0p_{1/2}^0i_{13/2}^{10})$ & 35.66\\
$ $ & {2}$^+$\hspace{0.5cm} & $\nu (f_{5/2}^0p_{3/2}^0p_{1/2}^0i_{13/2}^{10})$  & 37.46\\
$ $ & {4}$^+$\hspace{0.5cm} & $\nu (f_{5/2}^0p_{3/2}^0p_{1/2}^0i_{13/2}^{10})$  & 46.94\\
$ $ & {6}$^+$\hspace{0.5cm} & $\nu (f_{5/2}^0p_{3/2}^0p_{1/2}^0i_{13/2}^{10})$  & 49.08\\
$ $ & {8}$^+$\hspace{0.5cm} & $\nu (f_{5/2}^0p_{3/2}^0p_{1/2}^0i_{13/2}^{10})$  & 49.13\\

\hline

$^{194}$Pb &{0}$^+$\hspace{0.5cm} & $\nu (f_{5/2}^0p_{3/2}^0p_{1/2}^0i_{13/2}^{12})$ & 26.96\\
$ $  &{2}$^+$\hspace{0.5cm} & $\nu (f_{5/2}^0p_{3/2}^0p_{1/2}^0i_{13/2}^{12})$ & 25.27\\
$ $  &{4}$^+$\hspace{0.5cm} & $\nu (f_{5/2}^0p_{3/2}^0p_{1/2}^0i_{13/2}^{12})$ & 35.92\\
$ $  &{6}$^+$\hspace{0.5cm} & $\nu (f_{5/2}^0p_{3/2}^0p_{1/2}^0i_{13/2}^{12})$ & 39.96\\
$ $  &{8}$^+$\hspace{0.5cm} & $\nu (f_{5/2}^0p_{3/2}^0p_{1/2}^0i_{13/2}^{12})$ & 41.10\\

\hline 

\end{tabular}
\end{table}

\begin{figure}
\begin{center}
\includegraphics[width=9.00cm,height=5.5cm]{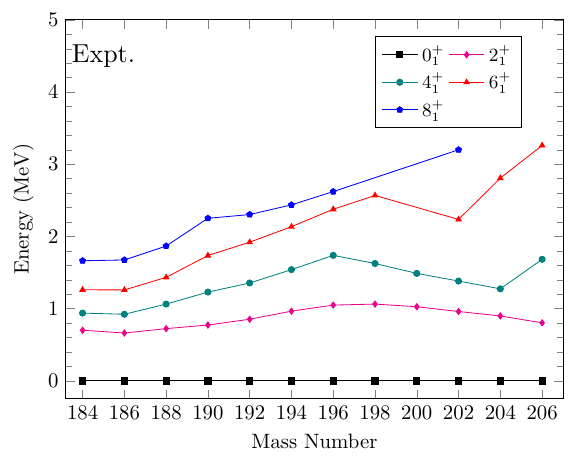}\\
\includegraphics[width=9.00cm,height=5.5cm]{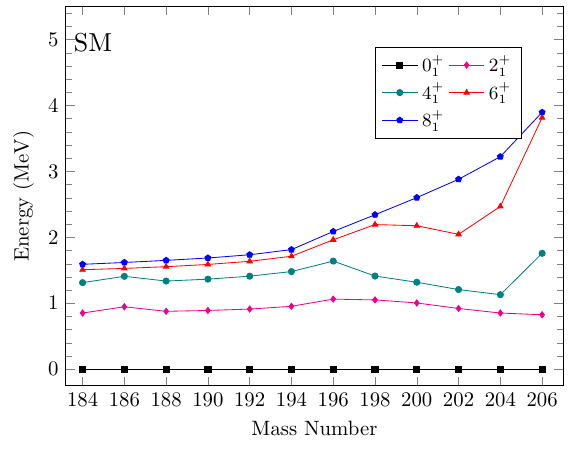}
\caption{\label{energy}Comparison between calculated and experimental \cite{NNDC} energy levels for $^{184-206}$Pb isotopes using KHH7B interaction.}
\end{center}
\end{figure}

\begin{figure}
\begin{center}
\includegraphics[width=9.00cm,height=5.0cm]{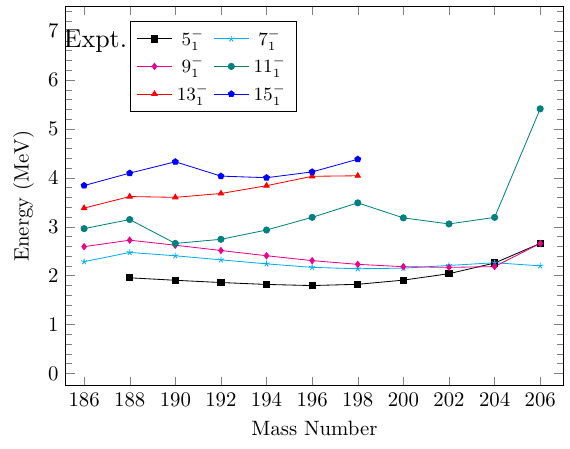}\\
\includegraphics[width=9.00cm,height=5.0cm]{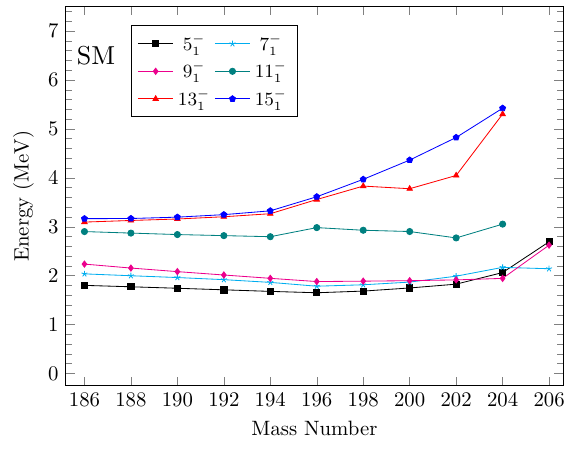}
\caption{\label{energy_negative}Comparison between calculated and experimental \cite{NNDC} energy levels for $^{186-206}$Pb isotopes using KHH7B interaction.}
\end{center}
\end{figure}

Fig. \ref{energy}, shows energy systematics of positive parity $0^+_1$, $2^+_1$, $4^+_1$, $6^+_1$, and $8^+_1$ states for $^{184-206}$Pb isotope, among which all the states up to $^{194}$Pb are obtained from $\nu(i_{13/2}^n)$, where $n$ is number of valence neutrons. Beyond $^{194}$Pb, up to $^{196-206}$Pb the states are obtained from the contribution of $p_{1/2}$, $p_{3/2}$, and $f_{5/2}$ orbitals, in addition with the $i_{13/2}$ orbital.

Fig. \ref{energy_negative} shows energy systematics of yrast odd $5^--15^-$ states for the $^{186-194}$Pb isotopes. Both the figures corresponding to experimental and theoretical data follows similar trend.
The $5^-$ state up to $^{196}$Pb and $7^-$ state up to $^{192}$Pb are obtained from the $\nu p_{3/2}^1i_{13/2}^{n-1}$ configurations, where n is the number of valance neutrons. The $9^-$ states up to $^{196}$Pb, and $11^-$ states from $^{186}$Pb to $^{196}$Pb are obtained due to $\nu f_{5/2}^1i_{13/2}^{n-1}$ configurations.

\begin{figure}
\begin{center}
\includegraphics[width=9.00cm,height=5.5cm]{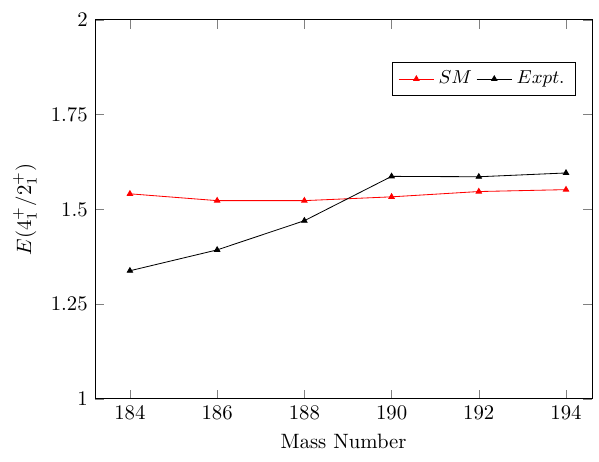}
\caption{\label{ratio}Comparison between calculated and experimental \cite{NNDC} E($4_1^+/2_1^+$) ratio for $^{184-194}$Pb isotopes using KHH7B interaction.}
\end{center}
\end{figure}

\begin{figure}
\begin{center}
\includegraphics[width=9.00cm,height=5.5cm]{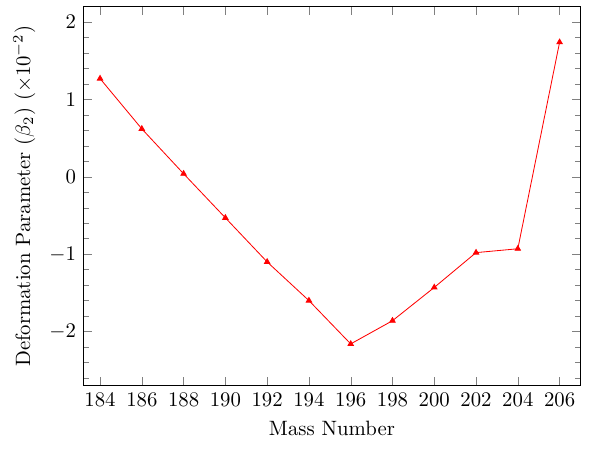}
\caption{\label{deformation}The calculated deformation parameter ($\beta_2$) for $^{184-206}$Pb isotope using KHH7B interaction.}
\end{center}
\end{figure}

We have calculated energy ratio $R_{4/2}$=$E(4^+_1)/E(2^+_1)$ for the $^{184-194}$Pb isotopes using KHH7B interaction corresponding to excitation energy of the first two excited states of ground state band which is almost constant (see Fig. \ref{ratio}) for whole chain and equal to $\approx$1.5, which shows these nuclei does not exhibit pure vibrational or rotational behavior, whereas they comes between spherical and deformed shapes i.e. these nuclei exhibits less collective behavior. The deviation between
the shell-model results and the experimental data in the neutron-deficient (mid-shell) region is observed.
We have also calculated a dimensionless quantity $R_{QB}$=$-Q(2^+_1)$/$\sqrt{B(E2;0^+_1 \rightarrow 2^+_1)}$ for the $^{184-194}$Pb using KHH7B interaction, both quadrupole moment and $B(E2)$ value corresponds to the deformation of nuclei. For these nuclei, $|R_{QB}|$ lies between 0.2 to 0.8. If $R_{QB}$ value is $<1$, then these nuclei are less deformed nuclei. In the rotational model of nuclei, spectroscopic quadrupole moment $Q_s$ and intrinsic quadrupole moment $Q_0$ are given by the relation $Q_s$=$\frac{3K^2-I(I+1)}{(I+1)(2I+3)}Q_0$, where K is the projection of the nuclear spin I on its symmetry axis. If we take assumption of $K=I$ then, $Q_s$=$\frac{I(2I-1)}{(I+1)(2I+3)}Q_0$, and we can write spectroscopic quadrupole moment in terms of deformation parameter ($\beta_2$) by using following equation \cite{qaudrupole}, 
\begin{equation}
    Q_s=\frac{3}{\sqrt{5\pi}}<r^2>Z\beta_2\frac{I(2I-1)}{(I+1)(2I+3)}, \end{equation} where $<r^2>$=1.44$\times A^{2/3}$ $fm^2$. We have calculated deformation parameter with the help of $Q(2^+_1)$ for the $^{184-206}$Pb isotopes, which is shown in Fig. \ref{deformation}. We obtain minimum deformation for the $^{184}$Pb for i.e. mid-shell Pb nuclei having $N$=102. Maximum deformation is obtained for $^{196}$Pb nucleus. The deformation parameter ($\beta_2$) calculated for $^{190-204}$Pb is negative therefore these nuclei exhibit oblate shape, while $\beta_2$ is positive for the rest of the nuclei shown in Fig. \ref{deformation}, therefore they exhibits prolate shape.

\begin{table*}
\centering
\caption{\label{be2} The calculated $B(E2)$ values (with KHH7B effective interaction) in units of W.u. for Pb isotopes compared to the experimental data (Expt.)
\cite{NNDC} corresponding to $e_p$ = 1.5$e$ and $e_n$ = 0.5$e$.  }

\vspace*{5mm}
\begin{tabular}{rrc|cccc}
\hline      
$^{184}$Pb  & Expt. & SM  & $^{186}$Pb & Expt. & SM  \\
& &    &   &     &    \\
\hline
${B(E2; J_i \rightarrow   J_f}$)   &     &   & ${B(E2; J_i \rightarrow   J_f}$) &  & \\
\hline
2$^+_1$ $\rightarrow$ 0$^+_1$   &  NA   & 0.6  & {2}$^+_1$ $\rightarrow$ {0}$^+_1$ &5.6(16) &1.0  \\
4$^+_1$ $\rightarrow$ 2$^+_1$   &  NA   & 0.8 &  {4}$^+_1$ $\rightarrow$ {2}$^+_1$&5.1$\times 10^2(12)$  & 0.3  \\
6$^+_1$ $\rightarrow$ 4$^+_1$   &   NA  & 0.7 & {6}$^+_1$ $\rightarrow$ {4}$^+_1$&4.5$\times 10^2(16)$ & 0.3  \\
8$^+_1$ $\rightarrow$ 6$^+_1$   &   NA  & 0.5 & {8}$^+_1$ $\rightarrow$ {6}$^+_1$&2.0$\times 10^2 (8)$ & 0.2  \\
10$^+_1$ $\rightarrow$ 8$^+_1$   &  NA   &   0.3 & {2}$^+_1$ $\rightarrow$ {0}$^+_2$& 190 (80)&2.2$\times 10^{-2}$\\
12$^+_1$ $\rightarrow$ 10$^+_1$   &  NA   & 0.1  & {10}$^+_1$ $\rightarrow$ {8}$^+_1$&NA &  0.1  \\
&   & & {12}$^+_1$ $\rightarrow$ {10}$^+_1$&NA & 4.3$\times10^{-2}$\\
&   & & {14}$^+_1$ $\rightarrow$ {12}$^+_1$ & NA& 0.7\\

\hline

\hline
& &    &   &     &    \\
$^{188}$Pb  & Expt. & SM  & $^{190}$Pb & Expt. & SM  \\
& &    &   &     &    \\
\hline
2$^+_1$ $\rightarrow$ 0$^+_1$   &  7(3)   & 1.3 & {2}$^+_1$ $\rightarrow$ {0}$^+_1$&NA &  1.5  \\
   4$^+_1$ $\rightarrow$ 2$^+_1$  &  163 (11)   &  0.1 & {8}$^+_1$ $\rightarrow$ {6}$^+_1$&NA &  $3.1\times 10^{-3}$  \\
   6$^+_1$ $\rightarrow$ 4$^+_1$  & 4.3$\times 10^2(7)$    &  0.1   & {9}$^-_1$ $\rightarrow$ {7}$^-_1$&NA & 0.7  \\
       8$^+_1$ $\rightarrow$ 6$^+_1$  & $3.3\times 10^2 (6)$    & 0.04    &  {10}$^+_1$ $\rightarrow$ {8}$^+_2$&0.012 &  0.8   \\
8$^-_1$ $\rightarrow$ 6$^-_1$  & $0.0120(20)$    &    0.3    &  {10}$^+_1$ $\rightarrow$ {8}$^+_1$& NA&$2.3\times 10^{-3}$\\      
12$^+_1$ $\rightarrow$ 10$^+_1$  & $0.0177 (15)$    &  $8.6\times 10^{-3}$   &  {12}$^+_1$ $\rightarrow$ {10}$^+_1$&NA &   $7.3\times 10^{-4}$  \\
19$^-_1$ $\rightarrow$ 17$^-_1$  & 0.031 (5)    &  1.3   &  {11}$^-_1$ $\rightarrow$ {9}$^-_1$&NA & 1.5   \\

\hline
& &    &   &     &    \\
$^{192}$Pb  & Expt. & SM  & $^{194}$Pb & Expt. & SM  \\
& &    &   &     &    \\
\hline
{2}$^+_1$ $\rightarrow$ {0}$^+_1$& NA& 1.6   &    {2}$^+_1$ $\rightarrow$ {0}$^+_1$&NA & 1.5   \\
{8}$^+_1$ $\rightarrow$ {6}$^+_1$&0.254(20)&  0.1  &    {9}$^-_1$ $\rightarrow$ {7}$^-_1$ &$2.15_{-32}^{+46}$ &0.3 \\
{9}$^-_1$ $\rightarrow$ {7}$^-_1$&6.1(6)&  0.5   &    {8}$^+_1$ $\rightarrow$ {6}$^+_1$ &$0.142_{-27}^{+44}$ &0.4 \\
{10}$^+_1$ $\rightarrow$ {8}$^+_1$&$0.9_{-9}^{+12}$& 0.1  &   {12}$^+_1$ $\rightarrow$ {10}$^+_1$ &$0.460_{-32}^{+34}$ &0.1  \\
 {10}$^+_1$ $\rightarrow$ {8}$^+_2$&0.003(3)& 0.1   &      {11}$^-_1$ $\rightarrow$ {9}$^-_1$ &${1.70\times 10^{-5}}_{-46}^{+44}$ & 1.2   \\
{12}$^+_1$ $\rightarrow$ {10}$^+_1$&0.16(3)& 0.02 &    {8}$^+_4$ $\rightarrow$ {6}$^+_4$ &${5.1\times 10^{2}}_{-17}^{+40}$ &0.1  \\
{11}$^-_1$ $\rightarrow$ {9}$^-_1$&0.00019(5)&  1.6   &    & &  \\
\hline

\hline
  \end{tabular}
 \end{table*}

\begin{table*}
\begin{center}
\caption{\label{bm1} The calculated $B(M1)$ values (with KHH7B effective interaction) in units of W.u. for Pb isotopes compared to the experimental data (Expt.)
\cite{NNDC,Dracoulis1} corresponding to $g_l^\nu$ = 0.00, $g_l^\pi$ = 1.00, and for spin angular momenta as $g_s^\nu$ = -3.826, $g_s^\pi$ = 5.585.  }

\vspace*{5mm}
\begin{tabular}{rrc|cccc}
\hline      
$^{184}$Pb  & Expt. & SM  & $^{186}$Pb & Expt. & SM  \\
& &    &   &     &    \\
\hline
${B(M1; J_i \rightarrow   J_f}$)   &     &   & ${B(M1; J_i \rightarrow   J_f}$) &  & \\
\hline
 6$^-_1$ $\rightarrow$ 5$^-_1$   &  NA   & 0.28 & {6}$^-_1$ $\rightarrow$ {5}$^-_1$ &NA &0.21 \\
 6$^-_1$ $\rightarrow$ 7$^-_1$   &  NA   & 0.04  &   {8}$^-_1$ $\rightarrow$ {7}$^-_1$&NA  & 0.45 \\
 8$^-_1$ $\rightarrow$ 7$^-_1$   &  NA   & 0.49  &   {6}$^-_1$ $\rightarrow$ {7}$^-_1$&NA & 0.04   \\
\hline

\hline
& &    &   &     &    \\
$^{188}$Pb  & Expt. & SM  & $^{190}$Pb & Expt. & SM  \\
& &    &   &     &    \\
\hline
 6$^-_1$ $\rightarrow$ 5$^-_1$   &  NA   & 0.14  & {6}$^-_1$ $\rightarrow$ {5}$^-_1$&NA &  0.09 \\
 6$^-_1$ $\rightarrow$ 7$^-_1$  &  NA   &  0.03  &  {6}$^-_1$ $\rightarrow$ {7}$^-_1$&NA & 0.03  \\
 8$^-_1$ $\rightarrow$ 7$^-_1$  & 9.1$\times10^{-7} (15)$    &  0.32 & {8}$^-_1$ $\rightarrow$ {7}$^-_1$&NA &  0.03  \\
{8}$^-_1$ $\rightarrow$ {9}$^-_1$& NA&0.01  &            {8}$^-_1$ $\rightarrow$ {9}$^-_1$& NA&0.09  \\

\hline
& &    &   &     &    \\
$^{192}$Pb  & Expt. & SM  & $^{194}$Pb & Expt. & SM  \\
& &    &   &     &    \\
\hline
  {6}$^-_1$ $\rightarrow$ {5}$^-_1$&NA&  0.06 & {6}$^-_1$ $\rightarrow$ {5}$^-_1$ &NA &0.04\\
{6}$^-_1$ $\rightarrow$ {7}$^-_1$&NA&  0.02  & {6}$^-_1$ $\rightarrow$ {7}$^-_1$ &NA &3.91$\times 10^{-3}$ \\
{8}$^-_1$ $\rightarrow$ {7}$^-_1$&NA& 0.01   & {8}$^-_1$ $\rightarrow$ {9}$^-_1$ &NA &0.16\\
{8}$^-_1$ $\rightarrow$ {9}$^-_1$&NA& 0.14  & {8}$^-_1$ $\rightarrow$ {7}$^-_1$ &NA & 0.07\\
& &    &    $8^+_4 \rightarrow 9^+_1$&$<3.3\times 10^{-6}$ & 5.58$\times 10^{-4}$\\
& &    &    $8^+_4 \rightarrow 8^+_1$&$11_{-3}^{+5}\times 10^{-7}$ & 1.12$\times 10^{-4}$\\
& &    &     $14^+_4 \rightarrow 13^+_1$ & NA&5.58$\times 10^{-4}$ \\
\hline

   
\hline
  \end{tabular}
  \end{center}
 \end{table*}

\begin{table}  
\centering
\caption{\label{qm} Comparison between the theoretical (with KHH7B effective interaction) and experimental \cite{NNDC} magnetic dipole moments $\mu$ (in $\mu_N$) and electric quadrupole moments $Q$ (in $eb$) in Pb isotopes. We have taken the effective charges as $(e_p,e_n)=$ (1.5, 0.5)$e$.  The  gyromagnetic ratios for orbital angular momenta are taken as $g_l^\nu$ = 0.00, $g_l^\pi$ = 1.00, and for spin angular momenta as $g_s^\nu$ = $-3.826$, $g_s^\pi$ = 5.585.}
\vspace*{5mm}
\begin{tabular}{c  c c  c   c c c c}
\hline
   Nucleus    & $J^\pi$   & $\mu$ ($\mu_N$) &  SM   & $Q (eb) $   &  SM \\
\hline

$^{184}$Pb & $2^+_1$  &NA   & -0.58    &  NA           & 0.10  \\
           & $4^+_1$  &NA &  -1.16  &       NA     & 0.11  \\
            & $6^+_1$  & NA& -1.76   &        NA    & 0.06  \\
            
 $^{186}$Pb & $2^+_1$  &NA & -0.57   &  NA          & 0.05  \\
           & $4^+_1$  &   NA      &  -1.16  &     NA       & 0.07  \\
            & $6^+_1$  & NA& -1.76   &     NA       & 0.04  \\
            & $7^-_1$  & NA& -0.99   &      NA      & -0.15  \\
         & $9^-_1$  & NA& -0.60   &      NA      & -0.23  \\
           & $11^-_1$  & NA& -1.55   &      NA      & -0.16  \\

   $^{188}$Pb    & $4^+_1$  & NA& -1.16   &     NA       &0.03  \\
            & $12^+_1$  & -2.148(72)& -3.52   &      NA      & -0.08  \\
            & $3^-_1$  & NA& -0.87   &      NA      & -0.05  \\
             & $8^-_1$  & -0.297 (24)& -3.44   &      NA      & -0.13  \\
             & $11^-_1$  & +11.33 (33)& -1.42   &      NA      & -0.16  \\

 $^{190}$Pb & $2^+_1$  & NA&  -0.55  &   NA         & -0.04  \\
           & $4^+_1$  &NA &  -1.15 &      NA      & -0.02  \\
                      & $6^+_1$  &NA &  -1.76  &      NA      & -0.004  \\
                      & $5^-_1$  &NA &-0.38  &       NA     & -0.03  \\
                    & $7^-_1$  &NA &-1.15  &       NA     & -0.026  \\
                      & $9^-_1$  &NA &-0.62  &       NA     & -0.14  \\
           
 $^{192}$Pb  & $2^+_1$  &NA &  -0.53  &    NA        & -0.09  \\
            & $4^+_1$  &NA &  -1.14  &  NA          & -0.07  \\
          & $6^+_1$  &NA &  -1.75  &  NA          & -0.03  \\
             & $12^+_1$  &-2.076(24) & -3.51   &    0.32(4)        & 0.13  \\
& $5^-_1$  &NA &-0.41  &       NA     & 0.023  \\
& $7^-_1$  &NA &-1.36  &       NA     & 0.05  \\
& $11^-_1$  &NA &-1.21  &       2.9(3)     & -0.11  \\

$^{194}$Pb & $2^+_1$  & NA& -0.50   &NA            & -0.14  \\
 & $4^+_1$  &NA &  -1.12  &NA           & -0.11  \\
  & $12^+_1$  &-2.00(2) &  -3.50  &0.49(3)           & 0.24  \\
  & $5^-_1$  &NA &  -0.44  &      NA      &  0.08 \\
  & $7^-_1$  &NA &   -1.54 &      NA      & 0.14  \\
   & $9^-_1$  &-0.38(14) &   -0.60 &      NA      & -0.005  \\        
  & $11^-_1$  &+11.3(2) &   -1.16 &      3.6(4)      & -0.07  \\

\hline
         
  \end{tabular}
 \end{table}

\begin{table*}
\begin{center}
\caption{Dominant Configurations (with KHH7B interaction) of the isomers in the Pb isotopes with their seniority. In this table we have also shown the  contribution of different seniority to generate a particular state and highlighted the major one.}
\label{t_sen}

\vspace*{5mm}

\begin{tabular}{ccc}
\hline
$J^{\pi}$   & Wavefunction  & \hspace{-1.5cm} Seniority \\
\hline

$^{188}$Pb   &  & \\
       8$^-$  & $\nu(p_{3/2}^1i_{13/2}^{5}$) (67.91 \%)
       &  {\bf $v$=2} [$v$ = 2 (89.73 \%) + $v$ =4 (9.75 \%)  +$v$ =6 (0.51 \%) ] \\
       11$^-$       &$\nu(f_{5/2}^1i_{13/2}^{5}$) (64.32 \%)  & {\bf $v$=4} [$v$ = 4 (78.24 \%) + $v$ =6 (21.76 \%)]\\
          12$^+$    &$\nu(i_{13/2}^{6}$)  (70.93 \%)    &   {\bf $v$=2} [$v$ = 2 (99.57 \%) + $v$ =4 (0.27 \%)  + $v$ =6 (0.16 \%)]\\
           19$^-$     &$\nu(f_{5/2}^1i_{13/2}^{5}$)  (84.17 \%)  &    {\bf $v$=4} [$v$ = 4 (92.21 \%) + $v$ =6 (7.79 \%)]\\
      \hline
      $^{190}$Pb   &  & \\
         10$^+$  & $\nu(i_{13/2}^{8})$  (58.90 \%)   &  {\bf $v$=2}  [$v$ = 2 (99.00 \%) + $v$ =4 (0.608 \%) + $v$ =6 (0.380 \%) ]\\
         12$^+$        &$\nu(i_{13/2}^{8}$) (59.22 \%)    &    {\bf $v$=2}  [$v$ = 2 (98.99 \%) + $v$ =4 (0.663\%) + $v$ =6 (0.335 \%) ] \\
        11$^-$        &$\nu(f_{5/2}^1i_{13/2}^{7}$)   (65.96 \%)    &    {\bf $v$=4} [$v$ = 4 (75.56 \%) + $v$ =6 (20.33 \%) + $v$ =8 (3.11 \%) ] \\  
        16$^+$        &$\nu(i_{13/2}^{8}$)   (4.95 \%)     & {\bf $v$=4} [$v$ = 4 (97.65 \%) + $v$ =6 (2.11 \%) + $v$ =8 (0.236 \%) ]\\  
        \hline
          $^{192}$Pb   &  & \\
       10$^+$    &$\nu(i_{13/2}^{10}$)     (49.14 \%)  &  {\bf $v$=2}   [ $v$ = 2 (98.13 \%) + $v$ =4 (1.24 \%)   + $v$=6 (0.6 \%)] \\  
       12$^+$        &$\nu(i_{13/2}^{10}$) (49.09 \%)  &   {\bf $v$=2}   [ $v$ = 2 (97.88 \%) + $v$ =4 (1.55 \%) + $v$ =6 (0.55 \%)]\\
       11$^-$             &$\nu(f_{5/2}^1i_{13/2}^{9}$)    (62.46 \%)   &  {\bf $v$=4}   [ $v$ = 4 (85.64 \%) + $v$ =6 (13.50 \%)]\\ 
  \hline   
    $^{194}$Pb   &  & \\
        {9}$^-$     & $\nu(f_{5/2}^1i_{13/2}^{11}$)   (46.24 \%)   &   {\bf $v$=2} [ $v$ = 2 (93.25 \%) + $v$ =4 (5.98 \%) + $v$ =6 (0.7 \%)]\\
        $ $ {8}$^+$            & $\nu(i_{13/2}^{12}$)   (41.10 \%)    &   {\bf $v$=2} [ $v$ = 2 (96.15 \%) + $v$ =4 (2.55 \%) + $v$ =6 (1.26 \%)]\\
        {10}$^+$                 & $\nu(i_{13/2}^{12}$)   (41.32 \%)     &  {\bf  $v$=2} [ $v$ = 2 (96.38 \%) + $v$ =4 (2.33 \%) + $v$ =6 (1.26 \%)]\\
      12$^+$           &$\nu(i_{13/2}^{12}$)  (39.91 \%)    &     {\bf $v$=2} [ $v$ = 2 (95.46 \%) + $v$ =4 (3.36 \%) + $v$ =6 (1.15 \%)]\\
      11$^-$            &$\nu(f_{5/2}^1i_{13/2}^{11}$)   (56.20 \%)    &     {\bf $v$=4} [ $v$ = 4 (93.20 \%) + $v$ =6 (5.37 \%) + $v$ =8 (1.40 \%)]\\ 
 
      \hline    
\end{tabular}
\end{center}
\end{table*}

\begin{table*}
\begin{center}
\caption{The computed half-life of isomeric states (with KHH7B effective interaction) for Pb isotopes compared to the experimental data (Expt.) \cite{NNDC}.}
\label{t_hl}
\vspace*{5mm}

\begin{tabular}{cccccccc}
\hline
&   &  &   & & &  \\
Isotope  &  $J^{\pi}_i$ & {$J^{\pi}_f$} & $E_{\gamma}$ & $B(E\lambda)$  & $B(E\lambda)$($e^2$fm$^{2\lambda}$)& Expt. & SM \\
& & & (MeV)  &   &  &$T_{1/2}$ & $T_{1/2}$  \\
& & &  &   & &  &  \\
\hline

& & &  &   & & &   \\
$^{188}$Pb 
  & $12_1^+$ & {$10^+_1$} & 0.062& $B(E2)$ & 0.540  & 97 (8) ns &20.0 $\mu$s   \\
& $19^-_1$&$17^-_1$ & 0.232 &$B(E2)$   &83.0 & 0.44 (6) $\mu$s& 8.07 ns  \\
& & & & & & &\\

\hline

& &  &  &   & & &  \\
$^{190}$Pb  & $10_1^+$ & {$8^+_1$} &0.053 & $B(E2)$ &  0.150 & 150 ns &74.22 $\mu$s   \\
  & $11_1^-$ & {$8^+_1$} &1.139 & $B(E3)$ &0.063   & 7.2 (6) $\mu$s & 7.5 ms  \\
   &          & {$8^+_2$} &0.792  & $B(E3)$ &0.013   & & \\
& & &  &   & & &   \\

\hline

& &  &  &   & & &   \\
$^{192}$Pb 
  & $12_1^+$ & {$10^+_1$} &0.058 & $B(E2)$ & 1.347  &  1.09 (4) $\mu$s&  8.11 $\mu$s \\

& & &  &   & & &   \\

\hline

& & &  &   & & &  \\
$^{194}$Pb  & $9_1^-$ & {$7^-_1$} &0.083 & $B(E2)$ &19.40   & 17 (3) ns &  0.50 $\mu$s \\
\hline

\end{tabular}
\end{center}
\end{table*}

\subsection{Isomeric states}
In Pb region, several isomers have been observed experimentally. Therefore, the study of their configuration and half-lives obtained from SM will be very beneficial. The seniority quantum number ($v)$, configurations, and half-lives of several isomeric states for Pb isotopes are presented in Tables \ref{t_sen} and \ref{t_hl}. 
The internal conversion coefficients (ICC)  \cite{Bricc} were used in the calculation of half-lives.

In spherical nuclei, isomeric states may arise via the splitting of high-$J$ nucleon pairs around the magic number. The Pb ($Z=82$) isotopes that we have taken into account have neutron numbers close to the magic number $N=126$. Thus, it is feasible to explain isomers in terms of seniority quantum numbers. Seniority quantum number is denoted by $v$ and defined as number of unpaired neutron or proton not coupled in pair to angular momentum $J$=0. {\color{black}From the configurations obtained using SM, information regarding seniority can be extracted. 
 If an isomeric state decays to another state having the same seniority, this causes strong suppression in the $B(E2)$ transition probability.  This hindrance arises because there is no change in seniority. In such cases, we identify the isomeric state as a seniority isomer.}

{\color{black}}

\begin{figure}
    \centering
    \includegraphics[width=0.6\linewidth]{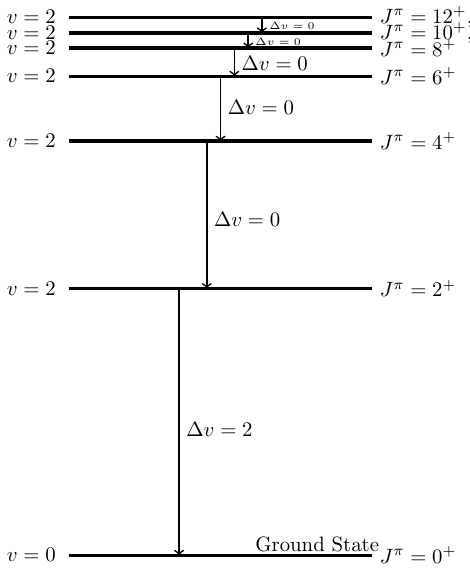}
    \caption{Seniority of yrast $0^+-12^+$ states of $^{194}$Pb.}
    \label{fig:seniority}
\end{figure}

\begin{figure}[h!]
    \centering
    \includegraphics[width=0.7\linewidth]{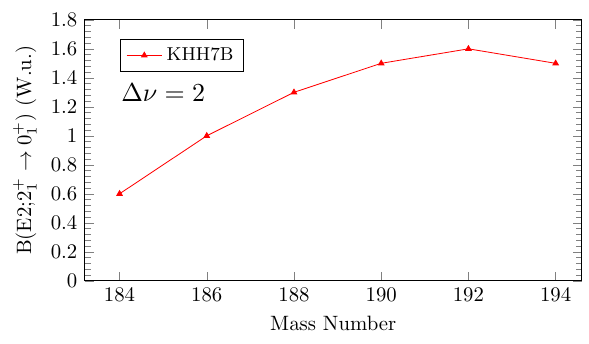}
    \includegraphics[width=0.7\linewidth]{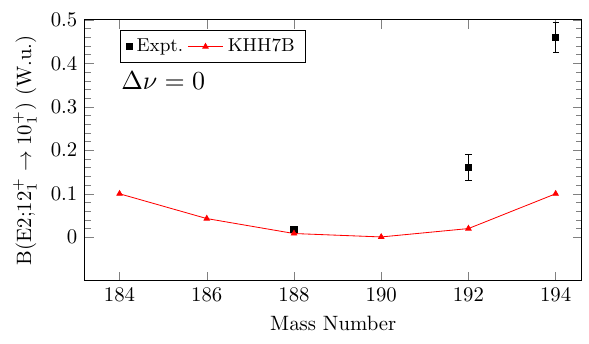}
    \includegraphics[width=0.7\linewidth]{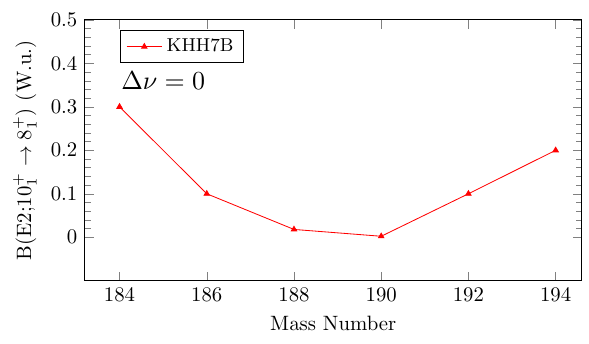}
    \caption{$B(E2)$ value for seniority changing transition ($2^+\rightarrow0^+$) ($\Delta \nu=2$) and seniority conserving transition ($12^+\rightarrow10^+$ and $10^+\rightarrow8^+$) ($\Delta \nu=0$). }
    \label{fig:be2_seniority}
\end{figure}

In the $^{188}$Pb isotope $8^-$, $11^-$, $12^+$, and ($19^-)$ are experimentally observed isomeric states, the SM configurations of these isomeric states are $8^-_1$ [$\nu(p_{3/2}^1i_{13/2}^5)$], $11^-_1$ [$\nu(p_{3/2}^1i_{13/2}^5)$], $12^+_1$ [$\nu(i_{13/2}^6)$], and $19^-_1$ [$\nu(f_{5/2}^1i_{13/2}^5)$] among which $8^-_1$ is obtained from one unpaired neutron in $p_{3/2}$, and another unpaired neutron in $i_{13/2}$ orbital i.e. $3/2^-\otimes13/2^+$. Whereas the $12^+_1$ isomeric state stems from one neutron pair breaking in $i_{13/2}$ orbital showing sphericity, therefore seniority ($v$) of these isomers are 2, the $12^+_1$ isomeric state is obtained in all even-even $^{188-194}$Pb isotopes discussed here. The $11^-_1$ and $19^-_1$ are coming from one unpaired neutron and one neutron pair breaking in $i_{13/2}$ orbital coupled with one unpaired neutron in $p_{3/2}$ orbital for $11^-$ and one unpaired neutron in $f_{5/2}$ orbital for $19^-_1$ isomeric state, therefore seniority of both of these isomers are 4. The $11^-_1$, and $12^+_1$ isomeric states decay via $E1$ and $E2$ transitions to the $10^+_1$ state, whereas $19^-_1$ isomer decays via $E2$ transition to the $17^-_1$ state. Here $10^+_1$, and $12^+_1$ state share similar configuration i.e. both of these states are coming from one neutron pair breaking in $i_{13/2}$ orbital which results even $2^+$-$12^+$ state, therefore both of these states possess same seniority quantum number due to which there is hindrance in their decay, which results in the very small theoretical $E2$ transition probability between these states i.e. 8.4$\times10^{-3}$ W.u., corresponding experimental $B(E2)$ value is also very small 0.0177(15) W.u., as reported in table \ref{be2}. We have calculated the half-lives of $12^+_1$ and $19^-_1$ isomers using $B(E2)$ values (from KHH7B interaction), the calculated values are 20.0 $\mu$s and 8.07 ns, whereas experimental values are 97 (8) ns and 0.44 (6) $\mu$s, respectively.

The experimentally observed isomers in the $^{190}$Pb isotope are $(10)^+$, $(12^+)$, $(11)^-$, and $(16^+)$, configuration obtained from our SM calculation for $10^+_1$ [$\nu(i_{13/2}^8)$], $12^+_1$ [$\nu(i_{13/2}^8)$], $11^-_1$ [$\nu(f_{5/2}^1i_{13/2}^7)$], and $16^+_1$ [$\nu(i_{13/2}^8)$] states are similar from both the interactions for these isomers. The $10^+_1$, and $12^+_1$ isomeric state is obtained from one neutron pair breaking in $i_{13/2}$ orbital, therefore seniority of these isomers are 2. Whereas, $11^-_1$ state is obtained from one unpaired neutron and one neutron pair breaking in $i_{13/2}$ orbital coupled with one unpaired neutron in $f_{5/2}$ orbital, and $16^+_1$ isomeric state is obtained from two neutron pair breaking in $i_{13/2}$ orbital, therefore seniority of these isomers are 4. Experimentally, the isomeric state $(10)^+$ decays via the $E2$ transition to the $(8)^+$ state, we have calculated half-life of this isomeric state from  KHH7B interaction with the help of $B(E2;10^+_1 \rightarrow 8^+_1$) value and the calculated value is 74.22 $\mu$s, whereas the experimental half-life is 150 ns. Experimentally, the $(11)^-$ isomeric state decays to $(8)^+$ state at 2.251, and 2.276 MeV excitation energies via $E3$ transitions, theoretically if we consider the configuration of $11^-$ state which is $\nu(f_{5/2}^1i_{13/2}^7)$, from the coupling of $f_{5/2}\otimes i_{13/2}^3$ we get minimum $J$ value equal to 4, therefore $E3$ transition between $11^-$, and $8^+$ state is not possible with this configuration, therefore {\color{black}we have to consider $1p-1h$ excitation across $N$ = $126$ }to calculate $B(E3;11^-_1\rightarrow8^+_1$), and $B(E3;11^-_1\rightarrow8^+_2$) values and the calculated value are 0.063, and 0.013 $e^2fm^6$, respectively. We have calculated the half-life of this isomeric state by using $B(E3;11^-_1\rightarrow8^+_1$) and $B(E3;11^-_1\rightarrow8^+_2$) values, which come to be 7.5 $ms$, whereas experimentally observed half-life is 7.2(6) $\mu s$, as shown in Table \ref{t_hl}.

The experimentally observed 
isomeric states in the $^{192}$Pb isotope are $(10)^+$, $(12^+)$, and $(11)^-$. The configurations obtained from the SM for the $10^+_1$ and $12^+_1$ states are $\nu(i_{13/2}^{10})$ i.e. these states are coming due to one neutron pair breaking in $i_{13/2}$ orbital resulting into seniority quantum number $v$ = 2, whereas configuration of the $11^-_1$ state is $\nu(f_{5/2}^1i_{13/2}^{9})$ i.e. this state is obtained from the coupling of one unpaired neutron in $f_{5/2}$, and one unpaired neutron and one neutron pair breaking in $i_{13/2}$ orbital resulting into seniority quantum number $v$ = 4. The $(12^+)$ state experimentally decays via $E2$ transition to the $(10)^+$ state, theoretically calculated half-life of this isomeric state using $B(E2;12^+_1 \rightarrow 10^+_1$) value is 8.11 $\mu$s. Corresponding experimental value is 1.09 (4) $\mu$s.

For the $^{194}$Pb, experimentally observed isomers are $(9)^-$, $(8)^+$, $(10)^+$, $(12^+)$, and $(11)^-$. The configurations obtained for these isomeric states from SM (using KHH7B) interaction are $9^-_1$ [$\nu(f_{5/2}^1i_{13/2}^{11})$], $8^+_1$ [$\nu(i_{13/2}^{12})$], $10^+_1$ [$\nu(i_{13/2}^{12})$], $12^+_1$ [$\nu(i_{13/2}^{12})$], and $11^-_1$ [$\nu(f_{5/2}^1i_{13/2}^{11})$]. Among these isomers $(9)^-$ decays via $E2$ transition for which our calculated (from KHH7B interaction) half-life using $B(E2;9^-_1 \rightarrow 7^-_1$) value is 0.50 $\mu s$ (corresponding experimental value is 17(3) ns). Our calculated $B(E2;8^+_4 \rightarrow 6^+_4$) value is very small in comparison to experimental value as reported in Table \ref{be2}, $8^+_4$ [$\nu(p_{3/2}^2i_{13/2}^{10})$] state is obtained from two neutron pair breaking in $i_{13/2}$ orbital, and $6^+_4$ [$\nu(f_{5/2}^1p_{3/2}^1i_{13/2}^{10})$] state is obtained from coupling of one neutrons in each $f_{5/2}$, and $p_{3/2}$ orbitals and one neutron pair breaking in $i_{13/2}$ orbital, therefore both of these states share same seniority quantum number which is equal to 4, hence obtained $B(E2)$ value is very small, due to hindrance in the decay of $8^+_4$ state to $6^+_4$ state.   
{\color{black}In $^{194}$Pb isotope $0^+-12^+$ states are obtained from the similar configuration $\nu(i_{13/2}^{12})$, i.e. $2^+-12^+$ states are formed by one neutron pair breaking in $i_{13/2}$ orbital, therefore corresponding seniority are 2 as shown in Fig. \ref{fig:seniority}, whereas $0^+$ state is obtained without pair breaking therefore its seniority is 0. In Fig. \ref{fig:be2_seniority}, $B(E2)$ values for seniority non-conserving ($\Delta \nu =2$) and seniority conserving ($\Delta \nu=0$) transitions are shown for $^{184-194}$Pb isotopes. For seniority changing transitions, we can see an inverted parabola, whereas for seniority conserving transitions, we obtain a parabola. The decay of $12^+$ state into $10^+$ state is hindered due to same seniority of both the states as we can see in Fig. \ref{fig:be2_seniority}, therefore, $12^+$ state is seniority isomer.}

\section{Summary and Conclusions}\label{sec4}

In the present work, we have done a systematic shell-model study of $^{184-194}$Pb isotopes using two different sets of effective interactions. The results corresponding to energy levels and electromagnetic properties are reported. We have also discussed shell-model results of isomeric states for these isotopes. 
The results of quadrupole and magnetic moments of different low-lying states are reported, where experimental data are not available. It will be quite useful 
to compare with the upcoming experimental data.

Further, we have drawn the following broad conclusions:
\begin{itemize}

\item  From the present study, it is observed that the shell-model study is quite challenging to reproduce the experimental data for positive parity states for $^{184-194}$Pb isotopes.

\item Present shell-model results are unable to reproduce the intruder $0_2^+$ in the same sequence as in the experiment. 

\item The sequence of negative parity states is nicely reproduced.

\item The ratio of $E(4_1^+/2_1^+)$ is almost constant around 1.5. These nuclei do not exhibit pure vibrational or rotational behavior. 

\item The shell-model results of the deformation parameter ($\beta_2$) is also reported.


\item The calculated magnetic moments of the $12_1^+$ state show good agreement with the available experimental data for the $^{188,192,194}$Pb isotopes.

\item {\color{black}The $11^-$ isomeric states in $^{188,190,192,194}$Pb are coming from dominant seniority $v$ = 4 (other contributions are 6, and 8), which decay to $9^-$ state of seniority $\nu$ = 2, due to the different seniority of both the states, $11^-$ is not a seniority isomer.
Whereas, $10^+$ isomeric states 
in $^{190,192,194}$Pb are having dominant seniority $v$ = 2 (other contributions are 4 and 6), which decays to $8^+$ state of seniority $\nu$ = 2, therefore $10^+$ isomer might be a seniority isomer.}

\item {\color{black}$12^+$ isomeric states in $^{188,190,192,194}$Pb are coming from dominant seniority $v$ = 2, which decays to $10^+$ isomeric state of the same seniority, due to which their decay is hindered. Therefore, the $12^+$ state is a seniority isomer in $^{188,190,192,194}$Pb isotopes.}

\item The shell-model result of the half-life of $12^+$ isomeric state in $^{192}$Pb is in good agreement with the experimental data and has seniority $v$ = 2.

\end{itemize}

\section*{Acknowledgments}
We acknowledge financial support from SERB (India),\\
CRG/2022/005167. We would like to thank the National Supercomputing Mission (NSM) for providing computing resources of ‘PARAM Ganga’ at the Indian Institute of Technology Roorkee. We also acknowledge useful discussions with Dr. Deepak Patel.

\end{document}